\documentclass[%
 aip,
 amsmath,amssymb,
 reprint,%
author-numerical,%
]{revtex4-1}
\usepackage{graphicx}
\usepackage{amsmath}
\usepackage[font=small,labelfont=bf,
   justification=justified,
   format=plain]{caption}
\usepackage{physics}
\usepackage{tcolorbox}
\usepackage{amsthm}
\usepackage{subcaption}
\usepackage{hyperref}
\usepackage{dcolumn}
\begin{document}

\title{Modelling Pulsed Deposition of Nanoparticles into films}

\author{Giacomo Becatti}
\affiliation{Department of Physics, University of Milan, Via Celoria, 16 I-20133 Italy}
\author{Francesca Baletto}
%\email{giacomo.becatti@studenti.unimi.it}
\email{francesca.baletto@unimi.it}
\affiliation{Department of Physics, University of Milan, Via Celoria, 16 I-20133 Italy}

\date{\today}
\begin{abstract}
     We propose a numerical tool to mimic the pulsed deposition of nanoparticles, a technique used to fabricate thin films from the deposition of nanoparticles upon a substrate. We employ such tool under different initial conditions, in particular exploring the effect of depositing an heterogeneous/homogenenous sample of nanoparticles in terms of their morphology (size and shape). We monitor how changing the nature of the building block affects the porosity and roughness of the grown nanofilms. We found a strong dependence on the size of the nanoparticles, following, in the low size regime, a growth of the porosity following a power law. 
     
\end{abstract}
\maketitle 

\section{Introduction}

Self-organising techniques, as self-assembly, are quite  well-established tools to manufacture nanomaterials. Among them, nanofilms obtained by the assembly of individual nanoparticles into films or wires are getting more and more attention.\cite{Mirigliano2021ElectricalFilms, Pike2020AtomicNetworks, Nadalini2023EngineeringProcess} 

Indeed, a low-energy deposition of nanoparticles and clusters retains their individuality leading to a complex structure, characterized by a dense network of junctions and grain boundaries \cite{Mirigliano2021ANanojunctions} which makes them potential candidates in a range of applications from catalysis\cite{Sanzone2021ScalingApplication} 
%\textcolor{red}{Front. Chem. Sci. Eng. 2021, 15(6): 1360–1379
%https://doi.org/10.1007/s11705-021-2101-7, 
%Scaling up of cluster beam deposition technology for
%%catalysis application
%Giuseppe Sanzone, Jinlong Yin, Hailin Sun}
to strain sensing\cite{Correa-Duarte2007OpticalFilms} to more exotic fields such as components of neuromorphic circuits.\cite{Sebastian2019ComputationalNetworks, Mirigliano2019Non-ohmicThreshold, Kuncic2021NeuromorphicProcessing}
%\textcolor{red}{Neuromorphic nanowire networks: principles, progress and future prospects for neuro-inspired information processing
%Zdenka Kuncic and Tomonobu Nakayama
%Article: 1894234 | 2021
%https://doi.org/10.1080/23746149.2021.1894234 \\} 
The latter are in need as the the booming of artificial intelligence that requires faster and faster physical processing units as brain-like computation can offer. Metallic nanoparticles (MNPs), as gold (AuNPs), can be used as building blocks for memristors. Nanoparticle networks produced by gas-phase cluster deposition technology show non-ohmic electrical behaviour and reproducible resistive switching. Those characteristic make them optimal candidates for reservoir computing\cite{Borghi2022InfluenceFilms, Minnai2017FacileDevices, Gronenberg2024InBehavior, Wu2023NeuromorphicFormation, Mallinson2024ExperimentalNanoparticles}.

The fabrication of devices based on cluster-assembled Au films requires a deep atomistic understanding of the influence of the individual cluster morphology on the nanoscale film structure and then of its properties. 
There is a profound difference between nanofilms from subsequent atom deposition or cluster assembled films. 
When films are grown by deposition of atoms present polycristalline structures characterized by a number of defects on the nano to micro scale. The growth process itself is characterized by the initial formation of islands, the coalescence of these islands then leads to the formation of a mostly continuous, nonporous structure.
On the contrary, films assembled by cluster deposition are characterized by a growth dynamics which strongly depends on the mass distribution of the impacting clusters.\cite{Mirigliano2021ElectricalFilms}
The initial deposition phase, up to around 70\% coverage of the surface, is characterized by an initial formation of islands, not too dissimilar to those initially formed by atom deposition. During this stage, the growth process was observed to be preferentially in the \emph{xy} plane, with the mean value of the cluster height increasing only by around 50\% with the average radius of the islands increasing by a more significant 160\% in agreement with experiments Mirigliano et al. \cite{Mirigliano2021ElectricalFilms, Mirigliano2019Non-ohmicThreshold}. 
The evolution of the surface coverage was found \cite{Borghi2018GrowthRegime} to follow the Deposition Diffusion and Aggregation (DDA) model, a model proposed by Jensen et al. \cite{Jensen1994DepositionGrowth} which describes the growth of nanostructures.
After the 70\% coverage is achieved, the growth process was observed to switch from a DDA growth model to a 3D 2 + 1 ballistic growth regime. The further diffusion of clusters on the surface is impeded by presence of the previously deposited clusters, acting as pinning centers and preventing to reach the substrate. 
The grain size distribution was observed to nearly perfectly overlap the distribution in the submonolayer film, thus indicating that in 3D ballistic growth, no significant grain growth is observed.\cite{Borghi2018GrowthRegime, Mirigliano2021ElectricalFilms}

Computational modelling is a robust tool to provide atomistic insights on the growth process. recently, the literature offers numerical models to explain the complex electrical behaviour of nanoparticles networks,\cite{Mambretti2022DynamicalNanojunctions}, as well as molecular dynamics to study the stability of nanojunctions between clusters \cite{Wu2022MolecularNetworks, Tiberi_TBP} the effect of the substrate in forming nanolinks or synapses (bridge) \cite{Wu2022MolecularNetworks, Wu2023NeuromorphicFormation}
%\textcolor{red}{M. Tiberi and F. Baletto, under review}. 
Nonetheless, few or none simulate the formation of nanofilms by cluster deposition. Here we present a numerical workflow, based on molecular dynamics, to mimic the formation of Au-nanofilms with a thickness between 30-40 nm as it might occur in a pulsed cluster deposition. We expand the Benetti and coworkers' approach\cite{Benetti2017Bottom-UpFilms} to AuNPs of arbitrary morphology. We consider different and spread distribution of sizes and shapes, with the aim to better reflecting the distributions of nanoparticles as in the beam deposition experiments.\cite{Mirigliano2021ElectricalFilms, Mirigliano2021ANanojunctions} 
We analysed the differences in the growth of films assembled by deposition of clusters homogeneous in size and shape, homogeneous in size and heterogeneous in shape and heterogeneous both in size and in shape. We also took into account the effects of increasing the kinetic energy of the deposited nanoparticles. Finally, we studied the evolution of the assembled systems at room temperature fro at least 15 ns. We proposed a free characterisation tool \texttt{NaMac} for the structural analysis of the nanofilms.
Changing the shape of the clusters and using mixed morphology sets of NPs for the deposition gave insights in the structural properties dependence on the shape and structure of the primeval clusters. What was observed was a more ordered structure in films assembled using, as part of the set, NPs with a FCC geometry, as truncated octahedra.
We observe that the porosity and the thickness of the nanofilms depends on the energy at which clusters are deposited. Furthermore, the porosity exhibit a power-law dependence on the size of the deposited individual nanoparticles. We also proposed a strain analysis showing that atoms are compressed at the surface and around the pores but there are several core regions where atomic distances are elongated.

\section{Methods}
We focus on the modelling of nanofilms obtained from the pulsed deposition of Au-nanoparticles (AuNPs) and deposited upon a Au(111) substrate.
\begin{figure}[hp]
\includegraphics[width = 1\linewidth]{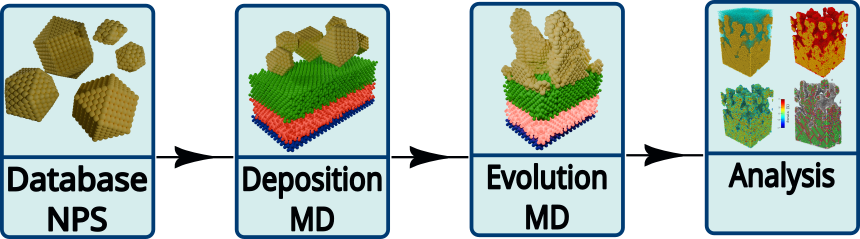}% Here is how to import EPS art
\caption{\label{fig:flow} Proposed numerical workflow for the virtual synthesis of cluster-assembled films. Each box represents of the stages. The database is a home-built collection of Au-clusters with different sizes and shapes while the analysis tool is an original code \texttt{NaMaC}, described in the text. The molecular dynamics engine is LAMMPS.}
\end{figure}

\begin{figure*}[htp]
\centering
\includegraphics[width = 1\linewidth]{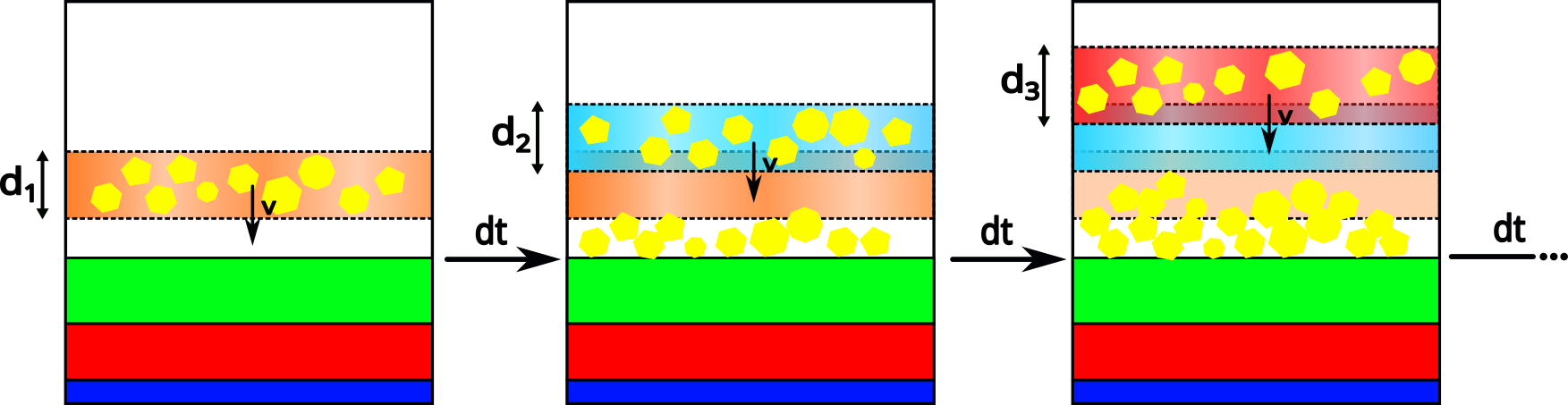}% Here is how to import EPS art
\caption{\label{fig:system_cfg} Depiction of the pulsed deposition model. The substrate is composed of three layers arranged from bottom to top as: the fixed layers to replicate bulk material (blue), the thermostated layers (red), and the free-to-move layers (green). The shaded regions are  the insertion boxes, coloured differently at each \emph{dt} interval. The yellow small polygons represent the $m$ AuNPs, randomly placed within the insertion box. They all possess a controllable kinetic energy directed towards the substrate, as labelled by the v-vector. The selected NPs are sourced from an existing database.}
\end{figure*}

Figure \ref{fig:flow} and \ref{fig:system_cfg} condensed the main objective of our model. Figure \ref{fig:flow} shows the overall workflow package of the proposed method, including the four different stages: (i) selection of NPs to be deposited from an existing database; (ii) deposition tool as discussed later; (iii) classical molecular dynamics simulation to check the evolution of the assembled nanosystem; (iv) original analysis of mechanical and structural properties of the nanofilms, using the software \texttt{NaMaC}, see next section.

Our molecular dynamics (MD) simulations utilize the LAMMPS software package \cite{Thompson2022LAMMPSScales}, leveraging the velocity Verlet algorithm \cite{Verlet1967ComputerMolecules} to solve the equations of motion with a timestep of 10 fs. This approach ensures a high degree of accuracy and enables long enough simulation of large Au systems (up to 10$^6$ atoms) \cite{Benetti2017Bottom-UpFilms}. For temperature control, we employ a Nosé-Hoover thermostat for 1 ns. The interatomic forces are derived using the second moment approximation of tight binding \cite{Gupta1981LatticeSurface}, a potential well-suited for describing the interactions among FCC atoms like Au. Further details on the parameters used are available in the input files in the supplementary information section. \textcolor{red}{Add citation}.

Figure \ref{fig:system_cfg} outlines the deposition tool, and highlights the main spatial parts in which we separate the system, namely the insertion (shadowed) and the substrate (the blue-red-green layers) regions.

The nanoparticles are selected from a database of structures, that includes AuNPs at different sizes shapes. Following observations that AuNPs can assume FCC-bulk geometries as well as non crystalline structures, the database contains geometrically build truncated octahedra (TOh) and cuboctahedra (COh), icosahedra (Ih) and decahedra (Dh). There are no limitation to the number and morphology of AuNPs in the database. In any event, before being used, the configuration is equilibrated at room temperature (300 K). Each nanoparticle is stored in the \textit{xyz} format. The reason is that this is the simplest configuration format. The workflow includes hence a converter to the \textit{molecule type} format used by LAMMPS. The size selection is done on the basis of the total number of atoms in the NP $N_{\alpha}$, and we consider "magic sizes" for Ih, Dh, TOh and COh.

Initially, AuNPs are randomly positioned in a so-called insertion region, shadowed areas in Figure \ref{fig:system_cfg}.
Insertion regions are generated every $dt$ time-interval, progressing along the $z$-axis with adjustable kinetic energy. %Periodic boundary conditions were applied in the \emph{xy} plane, while non-periodic fixed boundary conditions were maintained along the \emph{z} axis.  
%To give the substrate bulk like properties without periodic boundary conditions on the \emph{z} axis, we followed what was done by and divided the substrate in three layers, imposing on each layer a different constraint. 
The substrate, following the same idea as in Benetti et al. \cite{Benetti2017Bottom-UpFilms}, consists of three fixed monolayers, blue in Figure \ref{fig:system_cfg}, to represent the presence of a bulk;  a thermostated region (red) of 8 layers and at the topmost 10 free layers (green). The substrate is periodically repeated in $xy$, mimics a Au(111), and has an effective mass significantly greater than that of the deposited nanoparticles.

%The middle layer, represented in red in Figure \ref{fig:system_cfg} was thermostated with a Nosé Hoover thermostat to keep the temperature of the substrate constant at room temperature, the substrate without a heath bath would absorb the energy of the nanoparticles, which have sizes comparable with those of the substrate, causing as a result an increase of the temperature of the substrate at each impact. The topmost layer, in green in Figure \ref{fig:system_cfg} was instead kept free from any constraint, the point being to keep it free from non-physical artifacts, as the topmost layer is the most interested in the phenomena with the deposition of the nanoparticles and evolution of the assembled film. 

The insertion aims to reproduces simulating a pulsed deposition experiment, whereas AuNPs are deposited onto the substrate in time-separated pulses, such as in the SCBD (Supersonic Cluster Beam Deposition) experiments \cite{Mirigliano2021ElectricalFilms}. 
This is achieved by subdividing the total number of nanoparticles $M$ to be deposited into $j$-sets. Each set contains $m_j$-AuNPs randomly located in $j$-th insertion region and with a random orientation. The insertion box has a width of $d_j$ and moves with a velocity $v_j$ towards the substrate.
The system, substrate and the deposited NPs, evolve for a $dt$-time, after which another insertion box is created similar to the first one but shifted up along the deposition axis to preventing deposition upon existing clusters.

This insertion, deposition and translation of the insertion region is repeated until complete deposition of the desired number of nanoparticles, $M$.

We can tune the deposition time-interval (pulse) $\delta t$, the total number of $M$ nanoparticles and the total number of atoms in the nanofilm, $N_{tot} = \sum_{\alpha =1}^M N_{alpha} = \sum{j=1}^{insertions} \sum_{\alpha =1}^ {m_{j}} N_{alpha} $. The insertion region's width $d_j$, the number of nanoparticles per insertion box $m$, and their characteristics, namely their size $N_{alpha}$ and shape $\mathcal{S}_{\alpha}$, can be adjusted at each deposition. 
In the following we characterise the sample obtained using
$N_{tot} = 5*10^5 \pm 1000$ atoms, $\delta t = 200 ps$, 7 insertion, $d_j$ ranges between 220-320 \AA. We choose to study the different scenarios where we deposit
(i) monodispersed clusters: size and shape selected NPs; (ii) only size selected; (iii) fully heterogeneous samples or polydispersed clusters. We discuss these groups in details in the Results section.

%To the first group we consider nine situations where we deposit only Dh$_{80}$, Ih$_{147}$; Ih$_{309}$; Ih$_{561}$; TOh$_{976}$; Ih$_{1415}$; Ih$_{2057}$; Ih$_{3871}$; Ih$_{6525}$

\subsection{Characterization}
The characterization of the final assembled film was performed using a range of different codes.  We compiled a fast GPU accelerated Fortran code, Nanoporous Material Characterizer \texttt{NaMac}, to perform more specific investigations such as porosity evaluation, identification of surface atoms and atomic strain computation, which we will describe in the next sections. A basic structural identification of FCC and grain boundaries is done with Ovito \cite{Stukowski2010VisualizationTool}. within the film using the PTM (Polyhedral Template Matching) method \cite{Larsen2016RobustMatching}, using a RMSD cutoff of 1.2. The preference of this method over the CNA method was motivated by its higher robustness when faced with changing interatomic distances, due to high temperatures or, as in our case, more complex evolution of the morphology of the system.

\paragraph{NaMaC}:\texttt{NaMaC}, available on GitHub (available at \url{https://github.com/Giac97/NaMaC}), performs the computation of a series of different properties of interest in nanoporous materials. It computes the total porosity $\phi$ of the system, defined as the ratio between the empty and the total volume of the system:
\begin{equation}
    \phi = \frac{V_{\text{empty}}}{V_{\text{tot}}},
\end{equation}
a number, by its definition, defined between 0, for nonporous materials, to 1 for empty space.

Computation of the porosity is obtained by a Montecarlo integration, using a method that imitates evaluation of porosity using gas absorption \cite{Opletal2018PorosityPlus:Materials}. A certain number of test probe atoms is inserted at random location within the system, the insertion is then accepted according to:
\begin{equation}
    acc = \begin{cases}
        1 \quad \text{if}\quad d_{pj} > r_p + r_j\quad \forall j \\
        0 \quad \text{else}
    \end{cases},
\end{equation}
where $d_{pj}$ is the distance between the probe at its attempted insertion position and the \emph{j}-th atom of the system, while $r_p$ and $r_j$ are respectively the radius of the probe atom and of the \emph{j}-th atom. We repeat this check for all atoms in the systems, and, if no overlap between the probe atom and the film atoms is detected the test atom insertion is accepted, else is rejected. The porosity $\phi$ is then obtained as $N_{acc}/N_{tot}$, with $N_tot$ being the total number of attempted insertions. 

The possibility of defining a probe particle radius was implemented to give the possibility to the user to compare the results of the porosity evaluation for a simulated structure with the porosity in similar real systems, however, in this work, we opted for a more mathematical determination of the density using a point-like probe particle.

With \texttt{NaMaC}, aside from a determination of the overall porosity, it is also possible to determine a porosity profile by "slicing" the film in sections with the same thickness along the growth axis, the porosity is the determined using the method explained above for each of these slices. The result is a profile of the porosity with respect to the height inside the film, useful to study the degree of uniformity of the porosity inside the film and its behaviour as we progress with the deposits, in a similar way to the analysis performed by Benetti et al. \cite{Benetti2017Bottom-UpFilms}, albeit via a different method.

\begin{figure}[h]
\includegraphics[width = 1\linewidth]{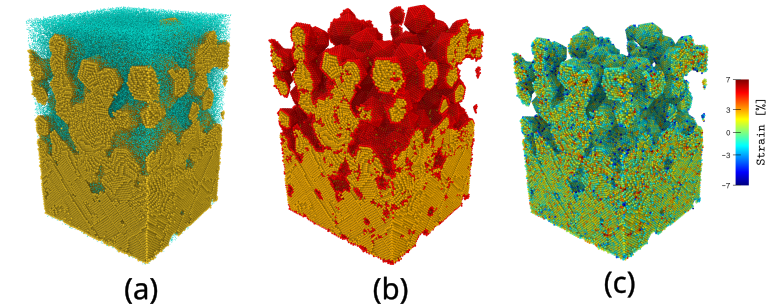}% Here is how to import EPS art
\caption{\label{fig:namac} Three outputs for different properties of the same system computed with \texttt{NaMaC}. (a) is the output from the computation of the porosity, the smaller light blue particles are the test particles, the yellow atoms are the gold atoms. (b) is the evaluation of the surface atoms, here colored in red, while (c) is from the computation of the strain, in a scale between -7\% (blue) to 7\% (red). }
\end{figure}

Furthermore, \texttt{NaMaC} can be used to determine, for each atom, whether it belongs to the surface. This identification is performed by computing, for each atom, its generalized coordination number (GCN) \cite{CalleVallejo2014FastNumbers, CalleVallejo2023TheElectrocatalysis}, an extension of the more conventional coordination number computed by normalizing the sum of the coordination numbers of each atom neighbours by a maximum coordination number specified based on the crystalline structure:
\begin{equation}
    GCN_i = \sum_{j \in n.n.}\frac{CN_j}{CN_{max}},
\end{equation}
in the case of gold we take $CN_{max} = 12$ as this is the coordination number found in FCC crystals. Each atom is then assigned as a surface atom if its GCN is less than a threshold value, assigned by test and inspection of the results, generally comprised between 9 and 10.

A further property computed by \texttt{NaMaC} is the atomic strain, computed following the definition by \cite{Nelli2023StrainNanoparticles}, where the strain is defined as the percentile ratio between the equilibrium position of the atoms in the ideal lattice and the position of the atoms in the systems. For each atom in the system the strain is computed via the following equation:
\begin{equation}
    s_i = 100\frac{1}{n_{b,i}}\sum_{j \in neigh(i)}\frac{d_{ij}-d_{bulk}}{d_{bulk}},
\end{equation}
where $n_{b,i}$ is the number of neighbours of the atom, the sum is over all neighbouring atoms, $d_{ij}$ is the distance between the atom and its j-th neighbour and $d_{bulk}$ is the equilibrium distance in a perfect crystal. 

From this definition we can then identify atoms with a positive strain to be on average compressed with respect to their ideal positions and decompressed if the computed strain is negative.
\section{Results}

We performed deposition simulation for three main different classes of systems. The first class we took into consideration was a rather ideal configuration, that is, for each deposition simulation, we used nanoparticles with the same morphology.

The following two sets of simulations included increasing spread of morphology dispersion, first by studying systems assembled by nanoparticles having the same geometrical shape and different sizes and then changing both sizes and shapes, getting closer to experimental conditions \cite{Mirigliano2021ElectricalFilms, Mirigliano2019Non-ohmicThreshold}.

We will present the results obtained for each initial condition in the following sections. For all of the simulations we kept the total number of atoms, rather than the number of cluster being deposited, constant in order to make a comparison with a similar sized crystalline nonporous bulk easier.

\subsection{Monodispersed Clusters}

We took into consideration cluster with sizes varying between 80 and 6525 atoms per clusters, this include nanoparticles with a diameter comprised between 1.08 and 4.76 nm.

The first characteristic we observed, which was qualitatively similar for all systems, was a dependency of the porosity profile between the surface of the substrate and the top of the film. 

\begin{figure}[h]
\includegraphics[width = 0.9\linewidth]{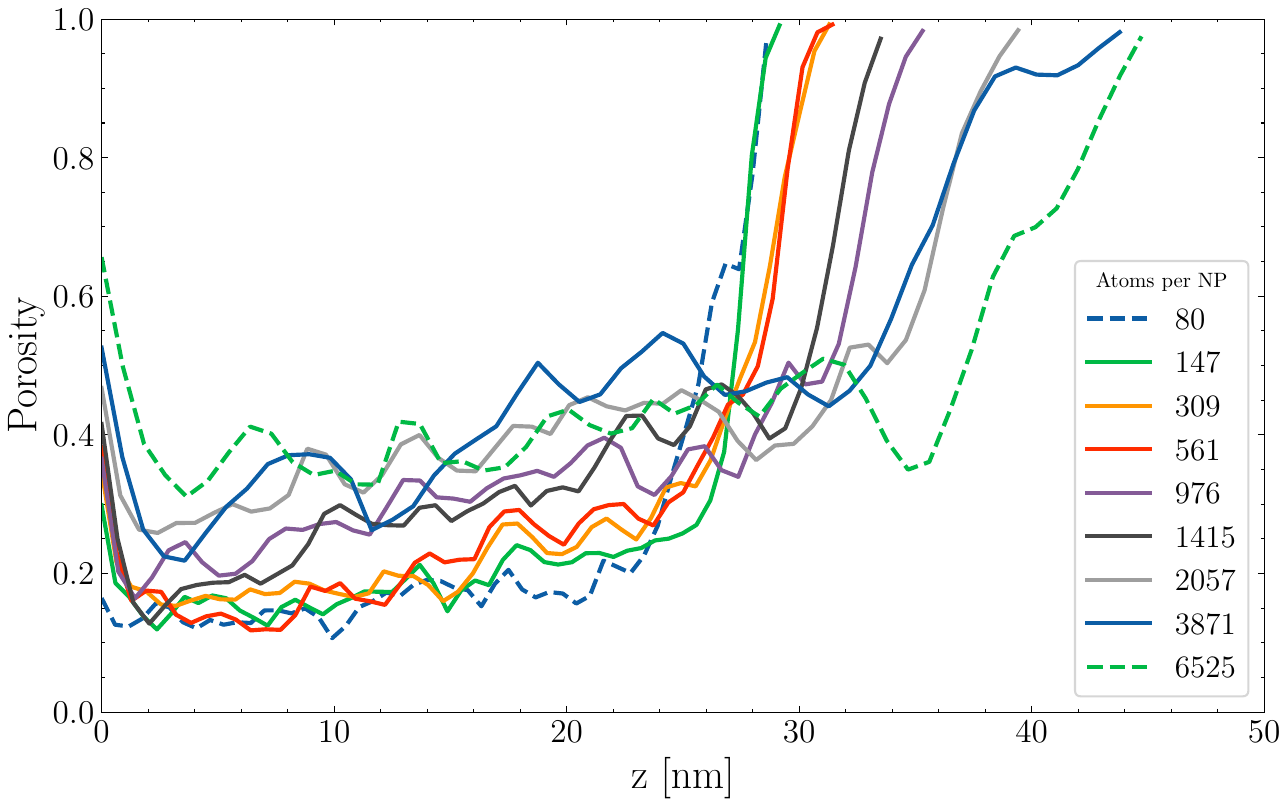}% Here is how to import EPS art
\caption{\label{fig:poro_profile_mono} Porosity vs height in the simulated films assembled by deposition of clusters with the same morphology.}
\end{figure}

What was observed was a first region, just above the substrate, characterized by a mostly uniform porosity, oscillating around a constant average value, then, 1 to 10 nm below the surface, this value depending on the size of the deposited nanoparticles, we observe a steep growth of the porosity.

This behaviour can be observed in fig. \ref{fig:poro_profile_mono}, where the profile of the porosity is displayed for all the systems taken into consideration. As mentioned we can observe a behaviour which is qualitatively similar in all systems, however, we can also observe how the initial average porosity increases with the size of the deposited clusters, varying between around 0.15 for the smallest up to 0.4 for the largest. 

We can also observe how the growth of the porosity below the surface follows a curve with decreasing steepness with the increase in cluster sizes, being almost vertical in systems assembled by smaller clusters, this being more evident in the film fabricated by deposition of icosahedral clusters composed of 147 atoms. This also influences the depth below the surface at which the porosity stops oscillating and starts growing.

\begin{figure}[h]
\includegraphics[width = 0.9\linewidth]{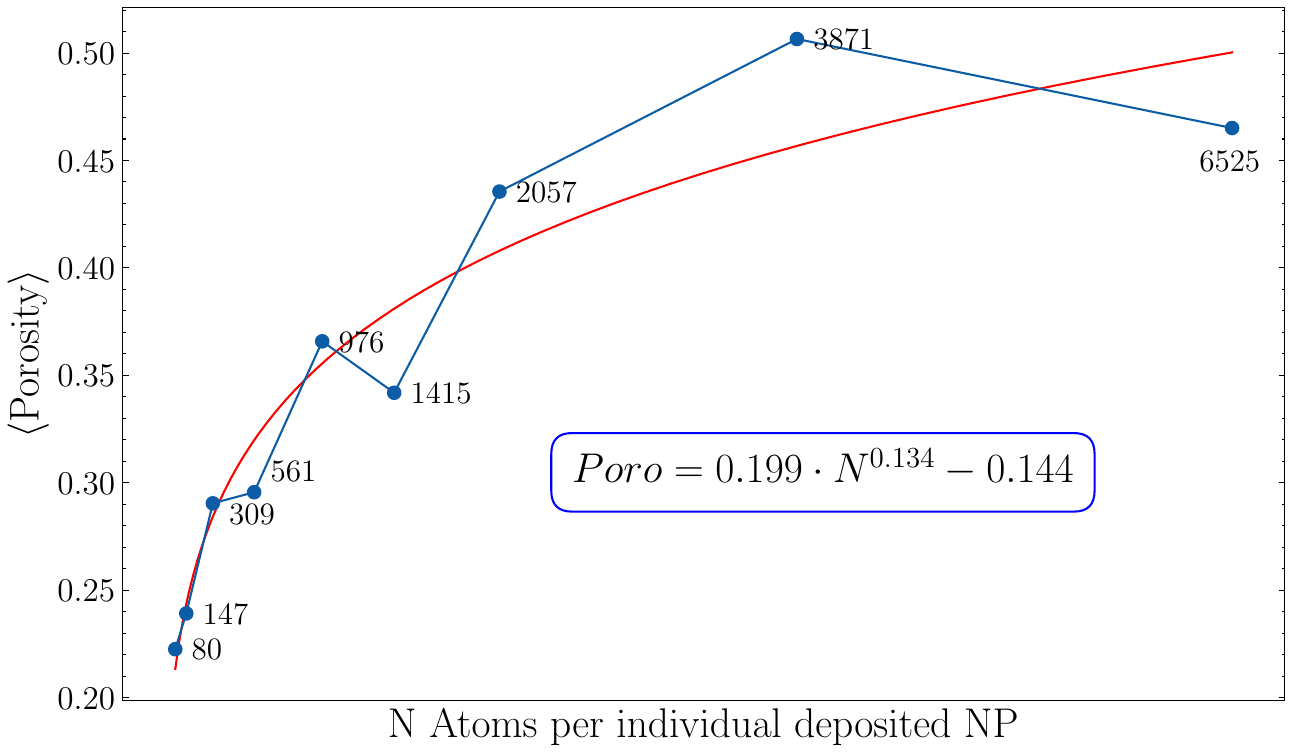}% Here is how to import EPS art
\caption{\label{fig:poro_growth}Average porosity fitted against the size of the individual nanoparticles deposited expressed in terms of the number of atoms per cluster.}
\end{figure}

A study of the average porosity, evaluated for the film as a whole, is displayed in figure \ref{fig:poro_growth}, where the average porosity was fitted against the size of the individual nanoparticles. 

We tried a simple power law fitting, which seems to give good results, this, however, is worth noting would be a good fit only valid in describing films with a relatively low porosity, as, by the way it is defined as a ratio between the empty and the total volume of a system, the porosity can only assume values between 0 and 1, so we can expect the actual curve of the average porosity to follow a more complicated law.

However simulating systems with even larger nanoparticles would require significantly larger systems, as, as the size of the cluster becomes comparable with the size of the system we can expect the porosity to start decreasing as a single nanoparticle might be able to fill the whole space.

\begin{figure}[h]
\includegraphics[width = 0.9\linewidth]{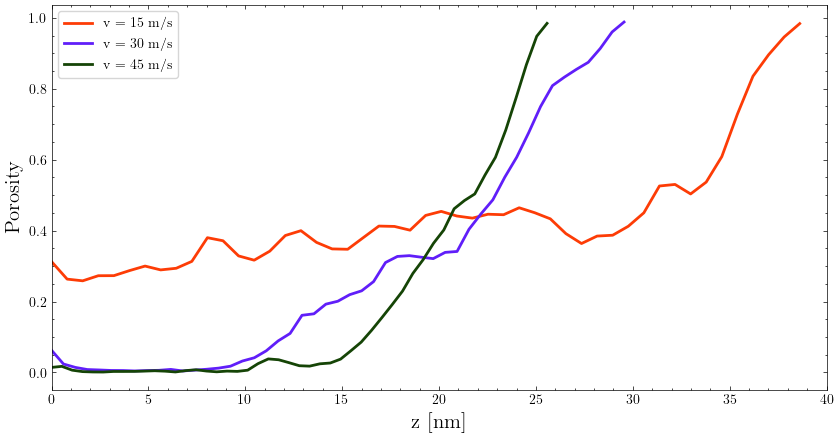}% Here is how to import EPS art
\caption{\label{fig:vel_comp_poro}Porosity profile observed in systems assembled by deposition of icosahedral clusters composed of 2056 atoms each at three different initial velocities.}
\end{figure}

For one of these systems we also tried to vary the kinetic energy assigned to the nanoparticles, to observe how this affects the structural properties of the assembled films. What we observed was a system that more closely resembles a polycrystalline structure, characterized be grains with different crystalline orientations. These structure were also characterized by a near zero porosity in the lower section of the film as reported in fig. \ref{fig:vel_comp_poro}.

\subsection{Size Selected Clusters}

After considering films assembled by deposition of clusters sharing the same morphology, we took into consideration four different sets of nanoparticles covering three different clusters each with the same size and different geometrical shape.  

\begin{table}[h]
\begin{ruledtabular}
  \begin{tabular}{ccccc}
   Set & $\text{d}_{avg}$ [nm]&  Shape 1& Shape 2 & Shape 3 \\
    \hline
    Set 1 & 1.4 & Au 147 Ih & Au 147 Co & Au 146 MDh \\
    Set 2 & 1.6 & Au 309 Ih & Au 309 Co & Au 318 MDh \\
    Set 3 & 2.0 & Au 561 Ih & Au 561 Co & Au 586 To \\
    Set 4 & 2.6 & Au 923 Ih & Au 976 To & Au 967 MDh \\
  \end{tabular}
    \caption{Set definition of sample obtained by size selected NPs.}
\label{tab:monosize}
\end{ruledtabular}
\end{table}

Table \ref{tab:monosize} reports the details of the four different sets taken into consideration, we took into consideration a mixture of geometries with internal crystalline and non crystalline symmetries.

\begin{figure}[h]
\includegraphics[width = 1\linewidth]{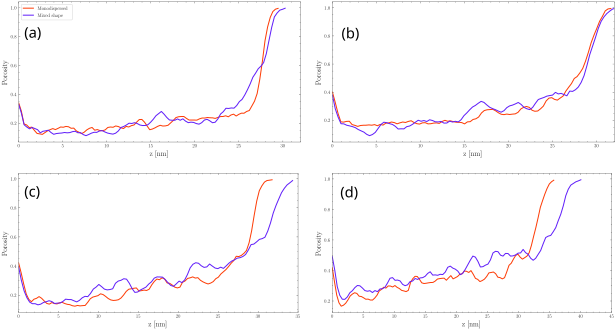}% Here is how to import EPS art
\caption{\label{fig:poro_monoshdisp}Comparison of the porosity profile between the systems assembled by nanoparticles with similar sizes and different shapes (blue) and similarly sized same morphology clusters (red). (a), (b), (c) and (d) correspond respectively to the nanoparticles with average diameter of 1.4, 1.6, 2.0 and 2.6 nm for the size selcted clusters and 147 Ih, 309 Ih, 561 Ih and 976 TO for the monodispersed clusters.}
\end{figure}

We found similar structural characteristic when comparing these systems with those assembled by similarly sized clusters, however, when looking at the porosity profile with the height and comparing it between the two systems (fig. \ref{fig:poro_monoshdisp}) we observe that, while the porosity does fluctuate around similar value in the lower section of the films, it starts to raise to one earlier and less smoothly in those systems presenting a dispersion in the geometrical shape of the nanoparticles.

\subsection{Polydispersed Clusters}

Real experiments involving cluster deposition by SCBD present a certain degree of dispersion of cluster morphology, what we aim to study here.
This is a tentative to be a more realistic in modelling film growth by considering a set of nanoparticles with a certain degree of dispersion in their morphology. In other words, we have in mind the deposition of individual nanoparticles which have different size and shape.

For this purpose we selected four sets with the first set larger cluster being also the second set smaller clusters and the same being repeated for the third set, with the clusters' sizes within the set being relatively close in size, plus a set with a larger spread of cluster sizes.

\begin{table}[h]
\begin{ruledtabular}
  \begin{tabular}{ccccc}
   Set & d [nm]&  NP 1& NP 2 & NP 3 \\
    \hline
    Set 1 & 1.32-1.70 & Au 147 Ih & Au 192 MDh & Au 309 Ih \\
    Set 2 & 1.70-2.40 & Au 309 Ih & Au 561 Ih  & Au 923 Ih \\
    Set 3 & 2.40-3.22 & Au 923 Ih & Au 1415 Ih & Au 2057 Ih \\
    Set 4 & 1.32-3.22 & Au 147 Ih & Au 923 Ih  & Au 2057 Ih \\
  \end{tabular}
    \caption{Set definition of sample obtained by NPs dispersed both in size and shape.}
\label{tab:polydisp}
\end{ruledtabular}
\end{table}

Table \ref{tab:polydisp} reports the four sets used in the assembly of films by simulating the deposition of clusters dispersed both in size and in shape, as with the previous simulations we keep constant, through all four sets, the total number of atoms, set to $5\cdot10^5$, the number of individual clusters is computed so that the total number of atoms is subdivided equally between all nanoparticles. 

It is still worth noting that such distribution do not really reflect the actual size and shape distribution found in real cluster deposition by SCBD experiments.

\begin{figure}[h]
\includegraphics[width = 1\linewidth]{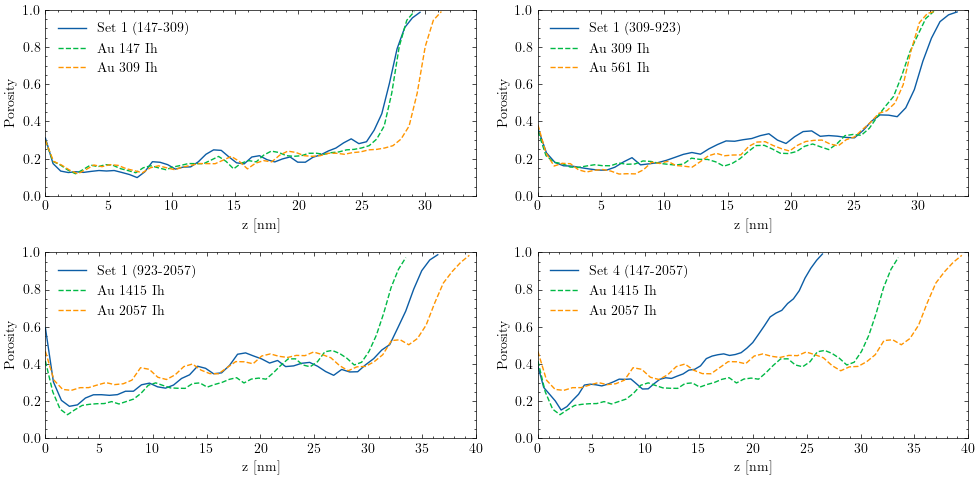}% Here is how to import EPS art
–\caption{\label{fig:poro_prof_disp} Porosity profile comparison between the films assembled by dispersed clusters and two similarly sized monodispersed clusters.}
\end{figure}

Figure \ref{fig:poro_prof_disp} shows a comparison between the porosity profile observed in films assembled by deposition of dispersed clusters and films assembled by monodispersed clusters.
We note that when the size distribution is between 1.3 and 3.2 nm the behaviour of porosity versus thickness is completely different from the other cases. 

Finally, defining the average thickness, $\mathcal{t}_{av}$ of the nanofilm when the porosity profile changes its slope. there is a clear indication that $\mathcal{t}_{av}$ depends on the size of the deposited clusters, increasing from 26 nm for the deposition of the smallest clusters up to 35 nm for the deposition of the largest Ih of more than 3 nm of diameter. Furthermore, increasing the size pf the deposited clusters we observe an almost flat region where the porosity oscillation around 0.25, then a gently increase up to 0.4, followed by the drastic increase for the topmost layer, where the porosity rockets to 1. The deposition of cluster with a size range distribution between 2.2$\pm$ 1 nm suggests that a reduced thickness of 19 nm and a porosity that smoothly approach 1 in 7 nm. We have too little statistics to comment in details on the size dependence of a reduction of the porosity as observed for the deposition of Ih$_{6025}$ around 36 nm of thickness.
The kinetic energy of the deposition affect both the porosity and the average thickness. At least it seems that exists a threshold above that the porosity falls to zero close to the substrate and decreasing the $\mathcal{t}_{av}$ down to 15 nm or less although at lower energy deposition the average thickness was the double if not more.

\subsection{Evolution of cluster-assembled films}

After completing the deposition of the film we let the system evolve for a total of 15 ns to observe the structural and thermodynamical evolution of the systems. What we observed was a similar behaviour, albeit on a different time scale, for all systems. 

Looking at the potential energy we observed an initial growth, followed by a plateau and a subsequent decrease, with the achievement of equilibrium only for systems assembled by small AuNPs with diameter in the 1.0 nm range. As the size grows we observed that this process takes place more slowly, not reaching equilibrium or even just reaching the plateau at the end of the growth of the potential energy in the systems assembled by nanoparticles with diameter above 2.8 nm.

\begin{figure*}[htp]
\centering
\includegraphics[width = 1\linewidth]{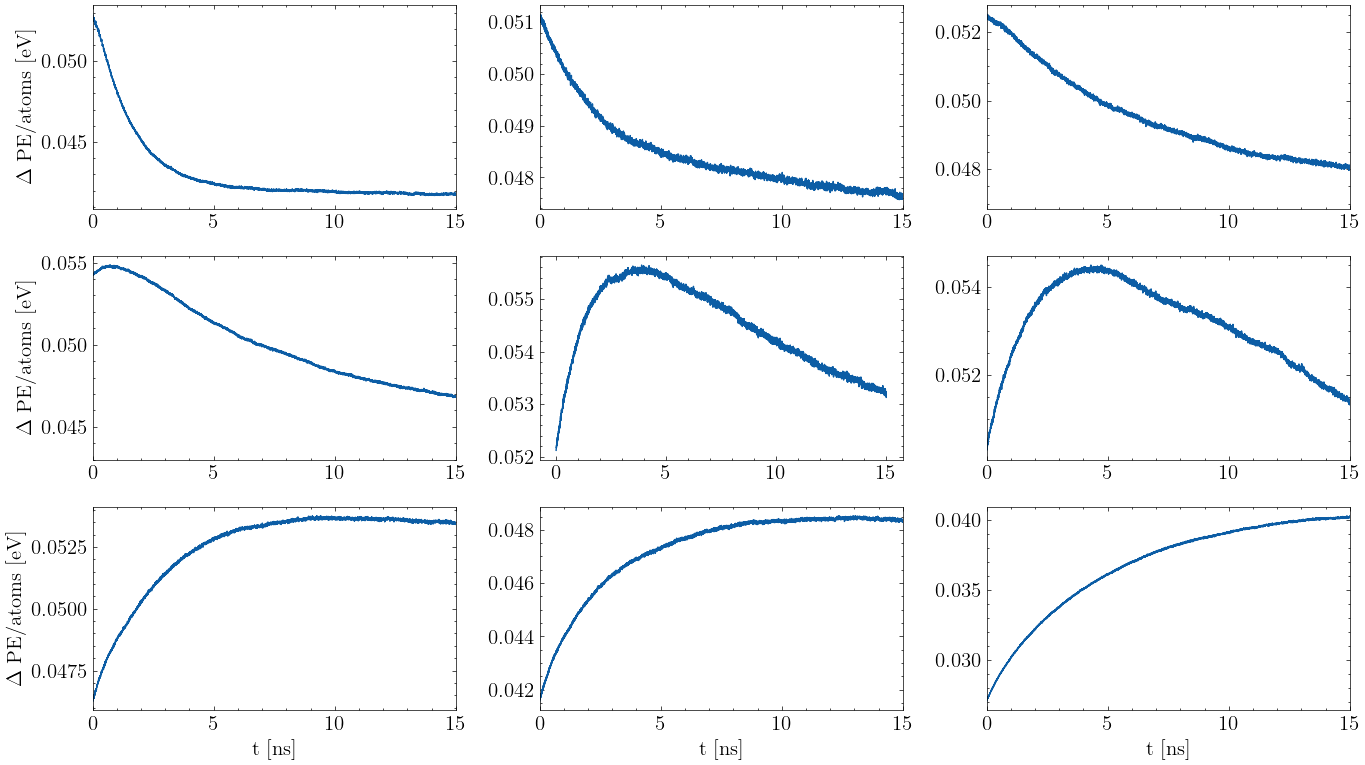}% Here is how to import EPS art
\caption{\label{fig:pe_monodisp} Potential energy evolution for systems assembled by clusters homogenous both in size and in shape. Left to right and top to bottom are the graphs for the films assembled by deposition of gold nanoparticles with the following number of atoms per cluster and shape: 80 mDh, 147 Ih, 309 Ih, 561 Ih, 976 TO, 1415 Ih, 2057 Ih, 3871 Ih, 6525 Ih.}
\end{figure*}

\begin{figure*}[htp]
\centering
\includegraphics[width = 1\linewidth]{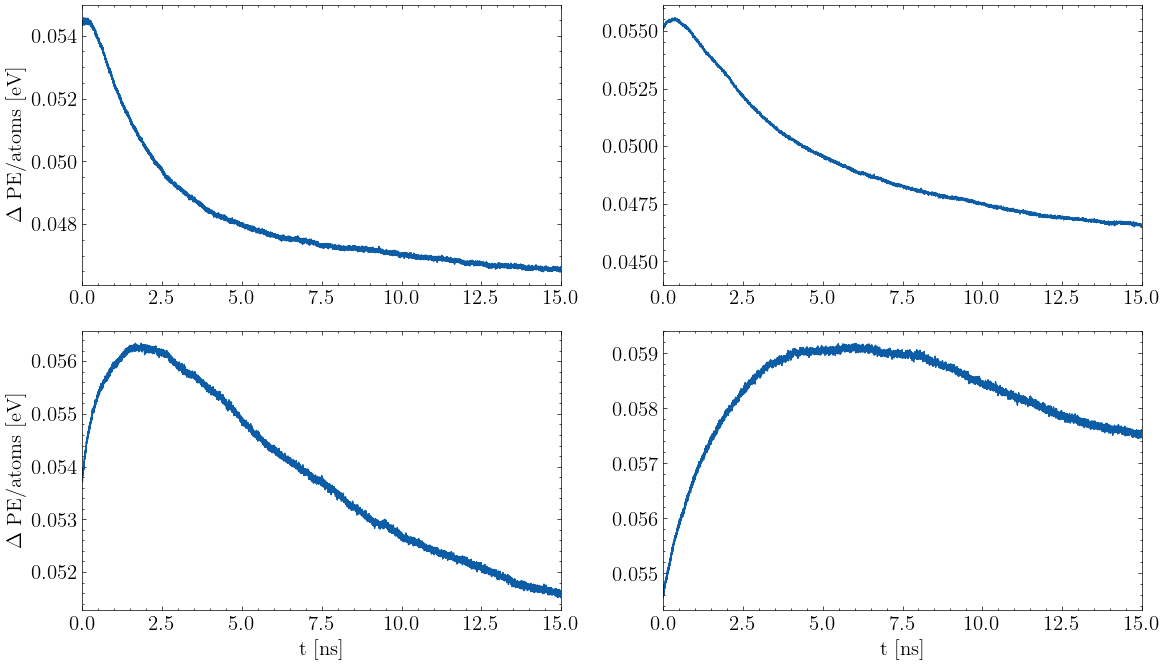}% Here is how to import EPS art
\caption{\label{fig:pe_shdips} Potential energy evolution for systems assembled by similarly sized clusters dispersed in shape.}
\end{figure*}

\begin{figure*}[htp]
\centering
\includegraphics[width = 1\linewidth]{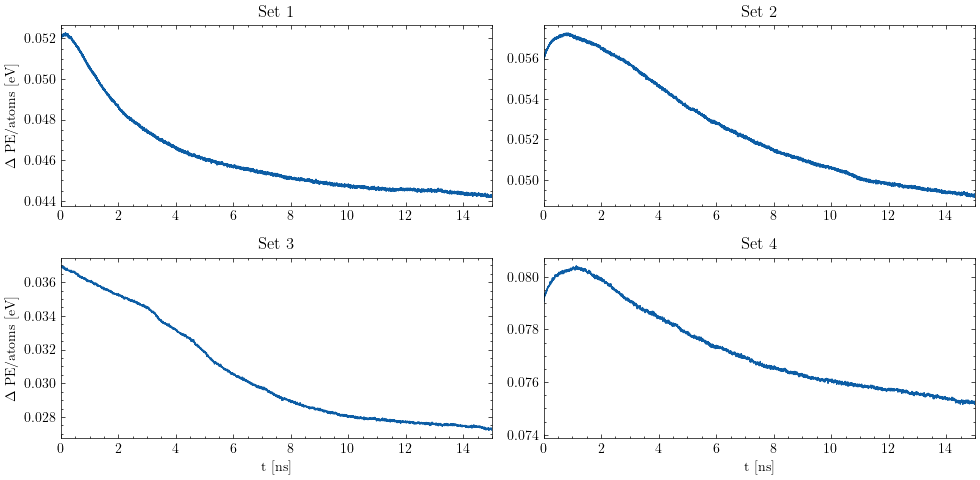}% Here is how to import EPS art
\caption{\label{fig:pe_polydisp} Potential energy evolution for systems assembled by similarly sized clusters dispersed in shape.}
\end{figure*}

When studying the evolution of the potential energy we compared it with the energy of a similarly sized slab of gold in order to evaluate the energetic cost of a more complex structure compared with an ordered crystalline surface. The evolution of the difference between the potential energy per atom in the cluster assembled film and the average per atom energy of the similarly sized slab is shown in figures \ref{fig:pe_monodisp} to \ref{fig:pe_polydisp}.

 During the time evolution of the nanofilms we note structural changes. What we observed, for all films, was a general decrease of the porosity and of the thickness of the films.

\begin{figure}[h]
\includegraphics[width = 0.9\linewidth]{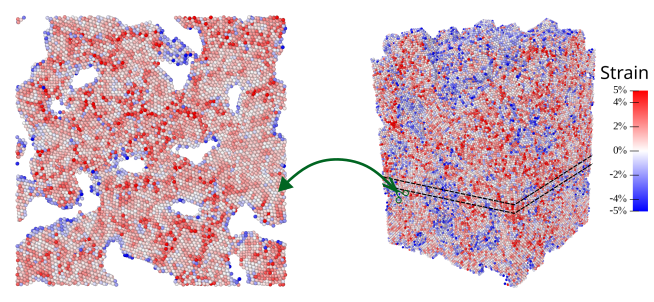}% Here is how to import EPS art
\caption{\label{fig:strain_ex}Strain in one of the cluster assembled films, atoms colored in shades of red (blue) have positive (negative) values of strain.}
\end{figure}

 PTM analysis also displayed an increase in the abundance of atoms identified as belonging to FCC structures, and, while the same analysis performed just after the completion of the deposition process displayed a greater relative abundance of FCC structures was significantly higher in films assembled by size selected clusters, due to the presence of crystalline clusters, with the progression of time we observed a generally more rapid increase in the abundance of FCC structures in those systems assembled by monodispersed clusters, leading to a reduction in the difference, going from 12.6\% at the end of the deposition to 6.4\% after the 15 ns evolution.

No significant variations were observed in the distribution of the strain within the films, changing by around 0.6\% for all systems throughout the time evolution. We observed for all systems a Gaussian like distribution of strain values, with a small positive average and a slight skew in the negative tail, with a higher average strain in systems assembled by non crystalline clusters. Relatively high value of strain (around 10\%) were observed in atoms belonging to vertexes which were not deformed upon impact of the nanoparticles, and, generally speaking we observed that atoms with positive strains were located within the film, while atoms having negative values of strain were located on the surface, see for example fig. \ref{fig:strain_ex}.

%Finally, To boost the use of the characterisation tool, we foreseen the inclusion of transport properties of the systems, both electrical\cite{Lopez-Suarez2021ModelingFilms, Lopez-Suarez2022MultiscaleFilms}, thermal and mechanical, e.g. elastic properites\cite{Benetti2017Bottom-UpFilms}.
\section{Conclusion}

Films and other structures assembled by nanoscale objects such as nanoparticles, present a wide range of interesting and peculiar properties differing from those found in bulk material of the same composition, which makes them suitable candidates for a number of different applications.

Molecular dynamics and numerical characterization methods were used to assemble and study the structural and thermodynamical properties of porous thin films using nanoparticles with different morphologies as building blocks. We first developed a model simulating a pulsed cluster deposition method, using this model we assembled several films with differing types and dipsersion of nanoparticles, subsequently we analysed the resulting virtual films and we let them evolve at a temperature of 300 K. The main findings from these simulations are:
\begin{enumerate}
    \item Depositing Au-NPs at low kinetic energies result in films with around 20-30\% of their volume being empty, this also leads to a higher surface to volume ratio compared with a film assembled by deposition of single atoms or high kinetic energy deposition of clusters. This porosity was observed to decrease during the evolution at finite temperature.
    \item Structurally these films present a high number of grains with different crystalline orientations, the size of these grains tends to increase with the evolution of the systems at finite temperature. Moreover, low energy deposition preserves the internal structure of the deposited clusters, resulting in a higher abundance of FCC structures in films where crystalline clusters were used as building blocks. 
    \item Apart from films assembled by small sub-nanometer clusters, we did not observe an equilibration of the potential energy over the course of the 15 ns time evolution.
\end{enumerate}

We hope that the proposed model provides a useful support for experiments and the fabrication devices employing the assembly of clusters as building blocks, as our tool can elucidate how the properties of the film can be optimised from the morphology of the deposited clusters.

\bibliography{references}

\newpage

\end{document}